\documentclass[a4paper,fleqn,preprint,times,12pt,oneside,onecolumn]{cas-sc}

\usepackage[authoryear]{natbib}
\usepackage{amsmath} 

\def\tsc#1{\csdef{#1}{\textsc{\lowercase{#1}}\xspace}}
\tsc{WGM}
\tsc{QE}

\usepackage{titlesec}
\titleformat{\section}
  {\normalfont\normalsize\bfseries}{\thesection.}{1em}{}
\titleformat{\subsection}
  {\normalfont\normalsize\itshape}{\thesubsection.}{1em}{}
\titleformat{\subsubsection}
  {\normalfont\normalsize\itshape}{\thesubsubsection.}{1em}{}

\usepackage{amsmath}
\usepackage{upgreek}
\usepackage{lineno}

\newcommand{\vect}[1]{\ensuremath{\mathbf{#1}}}

\begin{document}
\let\WriteBookmarks\relax
\def\floatpagepagefraction{1}
\def\textpagefraction{.001}

\shorttitle{}    

\shortauthors{JL Hesse-Withbroe \& KS Arquilla}  

\title [mode = title]{An Improved Rapid Performance Analysis Model for Solenoidal Magnetic Radiation Shields}  



%

\author{Joseph L. Hesse-Withbroe}[orcid=0000-0001-6812-5369]

\cormark[1]


\ead{joseph.hesse-withbroe@colorado.edu}


\credit{Conceptualization, data curation, formal analysis, investigation, methodology, project administration, resources, software, validation, visualization, writing --- original draft, writing --- review and editing}

\author{Katya S. Arquilla}


\ead{katya.arquilla@colorado.edu}


\credit{Conceptualization, project administration, supervision, writing --- review and editing}

\affiliation{organization={Smead Department of Aerospace Engineering Sciences, University of Colorado, Boulder},
            addressline={\\3775 Discovery Dr}, 
            city={Boulder},
            postcode={80303}, 
            state={CO},
            country={USA}}


\cortext[1]{Corresponding author}


\nonumnote{This material is based upon work supported by the National Science Foundation Graduate Research Fellowship Program under Grant No. DGE 2040434. Any opinions, findings, and conclusions or recommendations expressed in this material are those of the author and do not necessarily reflect the views of the National Science Foundation.}

\begin{abstract}
Astronauts participating in deep-space exploration missions will be exposed to significantly greater amounts of radiation than is typically encountered on Earth or in low Earth orbit (LEO), which poses significant risks to crew health and mission safety. Active magnetic radiation shields based on the Lorentz deflection of charged particles have the potential to reduce astronaut doses with lower mass costs than passive shielding techniques. Typically, active shielding performance is evaluated using high-fidelity Monte Carlo simulations, which are too computationally expensive to evaluate an entire trade space of shield designs. A rapid, semi-analytical model based on the High Charge and Energy Transport code (HZETRN) developed in 2014 provided an alternative method by which to evaluate the performance of solenoidal shields. However, various simplifying assumptions made in the original model have limited its accuracy, and therefore require evaluation and correction. In this work, a number of aspects of the original semi-analytical model are updated and validated by Monte Carlo simulation, then used to recharacterize the design trade space of solenoidal magnetic shields. The updated model predicts improved performance for weaker shields as compared to the original model, but greatly diminished performance for strong shields with bending powers greater than 20 T-m. Overall, the results indicate that magnetic shields enable significant mass savings over passive shields for mission scenarios where the requisite dose reduction is greater than about 60\% relative to free space, which includes most exploration missions longer than one year with significant time spent outside LEO.
\end{abstract}


\begin{highlights}
\item Crewed spaceflight
\item Space technology
\item Space radiation
\item Magnetic shields
\item GCR
\end{highlights}


\maketitle

\section{Introduction}\label{intro}

The deep space radiation environment, consisting of both solar energetic particles (SEPs) and galactic cosmic rays (GCRs), represents a major obstacle currently impeding our ability to conduct long-duration crewed exploration missions beyond LEO. Exposure to this space radiation causes myriad physiological effects that impact both crew and mission safety and are comprehensively reviewed in \citet{willey_individual_2021}. Chief among the  health detriments are acceleration of bone degradation \citep{macias_simulating_2016}, cataractogenesis \citep{ainsbury_radiation_2009}, cognitive decrements \citep{davis_individual_2014,hadley_exposure_2016,jewell_exposure_2018}, degenerative cardiovascular and renal diseases \citep{patel_risk_2015,siew_cosmic_2024}, and carcinogenesis \citep{gilbert_ionising_2009,ozasa_japanese_2018}.

The multitude of detrimental physiological effects of radiation exposure necessitate mitigation to limit their impact on crew health and mission safety.  NASA's established career permissible exposure limit (PEL) for spaceflight-associated radiation dose is 600 mSv for all astronauts \citep{francisco_nasa_2023}. This limit was designed to limit the risk of exposure-induced death associated with fatal carcinogenesis to 3\% for all crewmember demographics, evaluated at the upper 95\% confidence bound \citep{cucinotta_radiation_2010}. The total dose for an 860-day conjunction-class crewed Mars mission during solar maximum (e.g. Mars Design Reference Architecture 5.0 \citep{simon_nasas_2017}) for a lightly shielded habitat has been estimated to be 1.01 Sv, far in excess of the NASA career PEL \citep{hassler_mars_2014}. Passive shields have been shown to be incapable of reducing doses to acceptable levels for long-duration exploration missions without excessive mass costs on the order of hundreds of tons \citep{norman_early_2017,case_astronaut_2021}.

Alternative, mass-efficient radiation mitigation approaches are necessary to bring anticipated dose rates in line with permissible limits. Shields based on the Lorentz deflection of charged particles using superconducting electromagnets have been studied since the beginning of the space program \citep{ferrone_review_2023}. The dose reduction capability of these shields is typically estimated using high-fidelity Monte Carlo simulations of particle interaction with matter and magnetic fields (e.g., \citet{battiston_active_2011,westover_magnet_2014,vuolo_monte_2016,calvelli_novel_2017,chesny_galactic_2020,ferrone_reducing_2021,al_zaman_shielding_2023}). While these approaches enable a highly accurate quantification of shielding performance, they require significant computational resources and are prohibitively expensive for use in a trade study wherein the influence of key design parameters on shield performance must be quantified in order to identify optimal designs. Previous efforts have also often neglected to include comprehensive system mass estimates, which prevents their benchmarking against existing solutions and limits the utility of their results.

In this article, a brief overview of the operating principles of a magnetic radiation shield is provided. Then, improvements to a semi-analytical shield mass and dose estimation model \citep{washburn_analytical-hzetrn_2014,washburn_active_2015} based on the deterministic radiation transport code HZETRN \citep{slaba_updated_2020} are discussed. First, a correction factor is added to rectify a performance bias introduced by the original model's assumption of field uniformity. Second, a refined analytical description of GCR flux into the protected habitat is derived and is shown via Monte Carlo simulation to more accurately predict the magnetically-attenuated GCR spectrum inside the shield. Finally, the updated attenuation model also enables the estimation of dose throughout the entire volume of the protected habitat, whereas the original model was limited only to dose estimates at points along the central axis of the habitat. These updates to the model yield more accurate performance estimation of solenoidal radiation shields while maintaining the computational efficiency of the original model.

\section{Theory}


The fundamental physics that describes the operating principles of a solenoidal magnetic radiation shield are described in the following subsections.

\subsection{Magnetic Deflection of Charged Particles}\label{subsec:lorentz}
The motion of a charged particle in a pure magnetic field is described by the Lorentz equation, given by

\begin{equation}
  \vect{F}_L = q \vect{V} \times \vect{B}
  \label{eqn:lorentz}
\end{equation}

where $q$ is the particle's charge, $\vect{V}$ is its velocity, and $\vect{B}$ is the magnetic flux density at the location of the particle. From this expression, it can be shown that particle motion in a magnetic field of uniform flux density is circular. This gyromotion has a characteristic gyroradius given by 

\begin{equation}
  r_g = \gamma \frac{R}{{B}} = \gamma \frac{m_0 V_\perp}{q{B}} 
  \label{eqn:larmor}
\end{equation}

where $\gamma$ is the relativistic Lorentz factor, $R$ is the particle's magnetic rigidity, $m_0$ is the particle's rest mass, and $V_\perp$ is the component of the particle's velocity perpendicular to the magnetic field \citep{griffiths_introduction_2013}. Eq.\ (\ref{eqn:larmor}) can be rearranged to express the particle's kinetic energy as a function of its gyroradius.

\begin{equation}
  K(r_g) = \sqrt{\left(\frac{r_g qBc}{\sin\theta}\right)^2 + \left(m_0c^2\right)^2} - m_0c^2
  \label{eqn:larmor_kinetic}
\end{equation}

where $\theta$ is the angle between the particle's velocity and the direction of the uniform magnetic field.

\subsection{Generation of Magnetic Fields}\label{subsubsec:biotsavart}
A shield's strength is best described in terms of its bending power (the path-integrated magnetic flux density, expressed in T-m). The requisite bending powers for a shield capable of deflecting meaningful amounts of GCR radiation necessitate the generation of extremely strong magnetic fields on the order of $\sim$1 T. Superconducting electromagnets have previously been shown to be the most viable means of producing sufficient fields without excessive mass penalties \citep{washburn_model_2013}. A current loop $C$ produces a magnetic field at a point $\vect{r}$ according to the Biot-Savart Law

\begin{equation}
  \vect{B}(\vect{r}) = \frac{\mu_0}{4\pi}\int_{C}\frac{I\vect{\mathrm{d}\ell}\times \vect{r}' }{\lvert \vect{r}' \rvert ^3}
  \label{eqn:biotsavart}
\end{equation}

where $\mu_0$ is the permeability constant, $Id\ell$ is an infinitessimal current element along loop $C$, and $\vect{r}'$ is the displacement vector from the current element to $\vect{r}$. Analytical solutions to Eq.\ (\ref{eqn:biotsavart}) exist for a few ideal geometries. For example, the magnetic field at the center of an ideal, long, thin-walled solenoid aligned with the $Z$ axis is given by 

\begin{equation}
  \vect{B} = \frac{\mu_0 N I}{\sqrt{(D^2+h_s^2)}}\hat{z}
  \label{eqn:solenoid_field}
\end{equation}

where  $N$ is the number of coil windings, $I$ is the coil current, $D$ is the coil diameter, and $h_s$ is the coil length \citep{iwasa_case_1995}. The spatial variation of the field within the bore of the solenoid is generally assumed to be negligible.

\subsection{Overview of the Original Analysis Model}
\begin{figure}
  \centering
   \includegraphics[width=\textwidth]{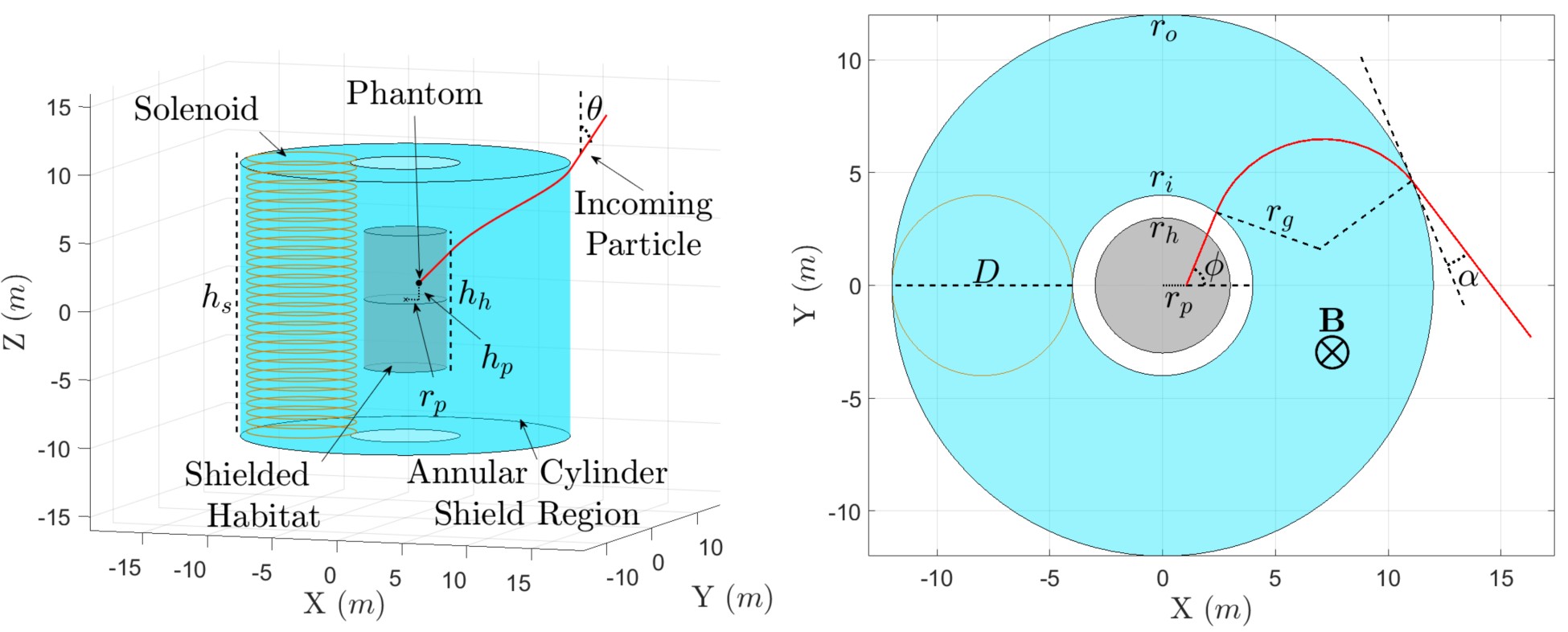}
    \caption{Geometry of a solenoidal shield surrounding a representative habitat (grey) with relevant features and dimensions labeled. The shield is modeled as an annular cylinder (blue) with uniform magnetic flux density $\vect{B}$ throughout. A GCR proton (red) approaching with polar incidence angle $\theta$ strikes the shield at a horizontal incidence angle $\alpha$ and is deflected along an arc of radius $r_g$. The particle possess sufficient energy to penetrate through the shield's inner wall and is assumed to travel in a straight line with an azimuth $\phi$ until it reaches a small, spherical water phantom located at position $\left(r=r_p,\phi=0, z=h_p \right)$ in a cylindrical coordinate system with origin at the center of the habitat. To maintain clarity, only one of the six solenoids in this design is shown.}\label{fig:shield_layout}
\end{figure}

The rapid analysis model discussed in later sections is an improved version of one first proposed by \citet{washburn_model_2013} (see also, \citet{washburn_analytical-hzetrn_2014,washburn_active_2015}). The model produces estimates of both the total mass of the shield and the shielding performance (i.e., dose reduction) with a runtime of approximately 20 s per design. This quick runtime enables the rapid evaluation of an entire trade space of solenoidal shield designs on the order of minutes to hours, depending on the number and resolution of the traded variables. To provide context for the improvements to the model, a general discussion of the modeling approach is included here. Interested readers are referred to the appendices of this article and to the original works for a more comprehensive description of the model. Key shield design and particle approach variables referenced throughout the following subsections are described and visualized to aid in clarity (Figure~\ref{fig:shield_layout}).

\subsubsection{Mass Estimation}
There are four primary subsystems that contribute significantly to the total mass of the shield. The superconductor subsystem is responsible for the production of the magnetic field, and is mainly composed of the superconducting cabling. The structure subsystem is responsible for maintaining the shape of the solenoids as they are powered on and their positioning with respect to the habitat and the other coils. The structural components must support tensile loads in the radial direction, compressive loads in the axial direction, and complex forces and torques resulting from the interaction of the magnetic fields of adjacent coils. The power subsystem supplies and transmits  electrical power to operate the fluxpumps that charge the coils and the cryocoolers that maintain their operating temperature below the superconducting critical current. The thermal subsystem rejects the heat loads incident on the coils, including solar and albedo heat loads, IR emissions from the habitat, and resistive heat loads from splices between adjacent superconducting cables. For each subsystem, basic physical principles and heuristic relationships are used to derive an estimate of that subsystem's mass. A representative example of such a derivation is provided in the appendix, while detailed derivations for all subsystems are available in \citet{washburn_model_2013} and \citet{washburn_active_2015}. 

\subsubsection{Dose Estimation}
To estimate the dose at a point inside the shield, the model must determine the influence on the GCR spectrum of both the magnetic attenuation afforded by the magnetic field and the passive shielding afforded by the protected habitat's pressure vessel and the physical shield materials (e.g., structural mass, superconducting cable mass). First, magnetic attenuation of the GCR spectrum is calculated as a function of particle energy for a point along the central axis of the habitat by

\begin{equation}
  \Phi_i^*(K) = 2\pi \Phi_i(K) \int_{0}^{\pi} H(K-K_{i,\text{co}}(\theta,\mathrm{r_{g}}(\theta))) \sin\theta d\theta
  \label{eqn:flux_attenuation}
\end{equation}

where $\Phi_i(K)$ is the unattenuated isotropic GCR flux of particle species $i$ at energy $K$ predicted by the Badhwar-O'Neill model, $H$ is the Heaviside step function, and $K_{i,\text{co}}(\theta,r_{g})$ is particle $i$'s cutoff kinetic energy (described by Eq.\ (\ref{eqn:larmor_kinetic})) as a function of that particle's polar approach angle $\theta$ (defined with respect to the $+Z$ axis) and its gyroradius. The form of the expression for the particle's gyroradius $r_g$ depends on the surface of the shield upon which the particle is incident (see Figure~\ref{fig:shield_regions}), which in turn depends on the polar approach angle $\theta$. Particles approaching near the poles of the shield (i.e., $\theta\approx$ or $\theta\approx\pi$) do not encounter the magnetic field prior to reaching the habitat, and so do not experience magnetic attenuation (i.e., $r_g=0$). Particles with relatively flat approaches centered around $\theta=\pi/2$ strike the shield along its vertical surface (the ``barrel'' surface). The gyroradius for these particles is given by 

\begin{equation}
  r_g = \frac{r_o^2 - r_i^2}{2r_o}
  \label{eqn:rg_washburn_barrel}
\end{equation}

where $r_o$ is the outer radius of the shield annulus and $r_i$ is the inner radius of the shield annulus. Approaching particles with intermediate polar angles between 0 and $\pi/2$ ($\pi/2$ and $\pi$) strike the shield along its upper (lower) horizontal surface (the ``outer transition'' surface). The gyroradius for these particles is given by 

\begin{equation}
  r_g(\theta) = \frac{\left( \left(h_s \pm 2h_p\right)\tan(\theta) \right)^2 - 4r_i^2}{4\left(h_s \pm 2h_p\right)\tan(\theta)}
  \label{eqn:rg_washburn_trans}
\end{equation}

where $h_p$ is the height along the central axis of the habitat at which the point dose is to be estimated. The $+$ case corresponds to particles entering through the bottom surface, while $-$ corresponds to the top surface. Eq.~(\ref{eqn:flux_attenuation}) is piecewise numerically integrated over the polar angle $\theta$ to produce a modified GCR flux spectrum representing the effect of the shield's magnetic attenuation on incoming radiation. The modified GCR spectrum is subsequently passed to the deterministic radiation transport code HZETRN \citep{slaba_faster_2010,slaba_updated_2020}. HZETRN is used to transport the magnetically-attenuated GCR spectrum through the mass of the shield materials located near the habitat and the mass of the habitat itself. Barrel and transition region particles are transported through $12\ \text{g}/\text{cm}^2$ of habitat aluminum, while endcap particles are transported through $25\ \text{g}/\text{cm}^2$. The transport procedure returns an estimate of the point dose equivalent in water produced by the actively- and passively-attenuated GCR spectrum, which is assumed to adequately represent the dose equivalent in tissue. In the original articles, HZETRN2010 was used to perform all transport calculations. With the release of HZETRN2020, the legacy 2010 codes were removed from NASA software repositories. Accordingly, all dose estimates presented in this article, for both the original and new models, were computed using HZETRN2020.

A number of assumptions underly the original model's derivation of magnetic GCR attenuation discussed above. First, an analytical expression for the kinetic energy cutoff $K_{i,\text{co}}$ is only achievable by assuming that the field magnitude is uniform throughout an annular cylindrical region, even though gaps between adjacent coils will have significantly reduced flux density. This results in an underprediction of total GCR flux inside the shield. Second, Eqs.~(\ref{eqn:rg_washburn_barrel}) and (\ref{eqn:rg_washburn_trans}) are only valid under the assumption that all particles approach along the most penetrating horizontal incidence angle (i.e. $\alpha=0$). This is a conservative assumption that overpredicts the amount of GCR flux inside the shield and positively biases the total estimated dose. Third, integrating over the approach angle $\theta$ prevents the inclusion of those particles whose initial velocities were not directed in such a way as to intersect the phantom, \textit{but, as a result of their interactions with the shield}, are ultimately redirected into the phantom. As will be shown in the following section, these particles can contribute significantly to the total dose. Finally, the derived equations are only valid for points located along the central axis of the habitat (i.e., $r_p=0$). As a conservative estimate, the original model computed the dose equivalent at the end of the habitat along the axial centerline (where dose is expected to be highest). While that convention is also used throughout the next section, a generalized formulation capable of computing the dose at any location within the habitat is also presented.

\section{Analysis Model Modernization}\label{sec:model}
In this section, the updates to Washburn's original performance evaluation model are discussed. Critical assumptions in the original model are reevaluated and generalized. The default parameters used to produce the data in this and subsequent sections are provided (Table~\ref{tab:defaults}). These default values are used unless otherwise specified.

\begin{table}
\caption{}\label{tab:defaults}
\begin{tabular*}{\tblwidth}{@{}LLL@{}}
\toprule
 Parameter & Value & Note \\ 
\midrule
 Solar Modulation Parameter ($\Phi$) & 481 MV & 1977 Solar Min (worst-case GCR conditions)\\
 Heavy Ion Fragmentation Model & RAADFRG & \citet{werneth_relativistic_2021} \\
 HZETRN Target Material & Water & Assumed to be representative of tissue dose \\
 Phantom Radial Location ($r_p$) & 0 & GCR flux integrals simpler along central axis\\
 Phantom Axial Location ($h_p$) & 4.95 m & End of habitat, worst-case position \\
 Habitat Height ($h_h$) & 10 m & Representative Mars transit habitat dimensions\\
 Habitat Radius ($r_h$) & 3 m &  \\
 Habitat Sidewall Thickness & 12 $\text{g}/\text{cm}^2$ Al & Same as original model \\
 Habitat Endcap Thickness & 25 $\text{g}/\text{cm}^2$ Al & Same as original model\\
 Shield Inner Radius ($r_i$) & 4 m & 1 m gap between habitat and shield \\

\bottomrule
\end{tabular*}
\end{table}

\subsection{Annular Cylinder Assumption}

A solenoidal radiation shield is formed by surrounding a space habitat with a number of solenoid coils (Figure~\ref{fig:shield_layout}). To simplify analysis, the original model modeled the shield as an annular cylinder with a uniform flux density given by Eq.\ (\ref{eqn:solenoid_field}), including in the gap regions between adjacent coils. While the assumption of field uniformity in the annular volume is necessary to enable the derivation of an analytical expression for GCR flux attenutation, in its original form this assumption positively biases the performance of the shields. In reality, the gaps between shield coils will have a substantially lower field flux density than that predicted by Eq.\ (\ref{eqn:solenoid_field}), and the average bending power of the shield will be lower than predicted by the model. To rectify the biases introduced by the averaging process while maintaining the field uniformity property, the field flux density within the coils is scaled by a geometrical factor describing the ratio of the volume of the annular cylinder to the volume contained within the coils

\begin{equation}
  \epsilon = \frac{h_s\pi\left(r_o^2-r_i^2\right)}{N_{\text{coils}} h_s \pi \left(\frac{D}{2}\right)^2} = \frac{r_o^2-r_i^2}{N_{\text{coils}}\left(\frac{D}{2}\right)^2}
  \label{eqn:B_scale_factor}
\end{equation}

where $N_{\text{coils}}$ is the number of solenoid coils that form the shield, and $r_o$ and $r_i$ are the outermost and innermost radii of the shield, respectively. To enable the comprehensive evaluation of the design trade space, the original model allowed $N_{\text{coils}}$ to vary continuously according to the interpolation function 

\begin{equation}
  N_{\text{coils}} = \frac{2\pi\left(r_i+\frac{D}{2}\right)}{D}
  \label{eqn:N_coils}
\end{equation}

Substituting Eq.\ (\ref{eqn:N_coils}) into equation (\ref{eqn:B_scale_factor}) and simplifying yields $\epsilon=4/\pi$. Accordingly, solenoids producing within their bores a field of $4 B/\pi$ will produce an averaged field of $B$ across the entire annular cylinder. The correction scale factor is applied only to the mass estimation component of the analysis model, whereas the dose estimation components remain unscaled. Applying this scaling to the mass estimation model is intended to yield a more accurate representation of the additional mass of shield components that would be required to produce a field that is approximable by an annular cylinder with uniform flux density $B$.

\subsection{Magnetic GCR Spectrum Attenuation}
Charged particles entering the shield region will be deflected via the Lorentz force (Eq.\ (\ref{eqn:lorentz})). The deflections caused by the magnetic field will naturally cause some particles that would have otherwise intersected the phantom to deflect away from the habitat, but will also cause other particles that would have originally missed the habitat to be ``focused'' into the habitat and intersect the phantom. An accurate description of the GCR environment within the protected habitat requires treatment of both cases. Whether a particle ultimately penetrates the shield depends on the size of the shield and its flux density, the particle's rigidity, the particle's direction of incidence (defined by the polar angle $\theta$ and azimuthal angle $\alpha$), and the shield surface upon which the particle is incident. Here, a brief overview of key improvements to the magnetic attenuation model and their effects is provided, while the full derivation of the expressions describing GCR flux attenuation is left to the appendix.

\begin{figure}
  \centering
   \includegraphics[width=\textwidth]{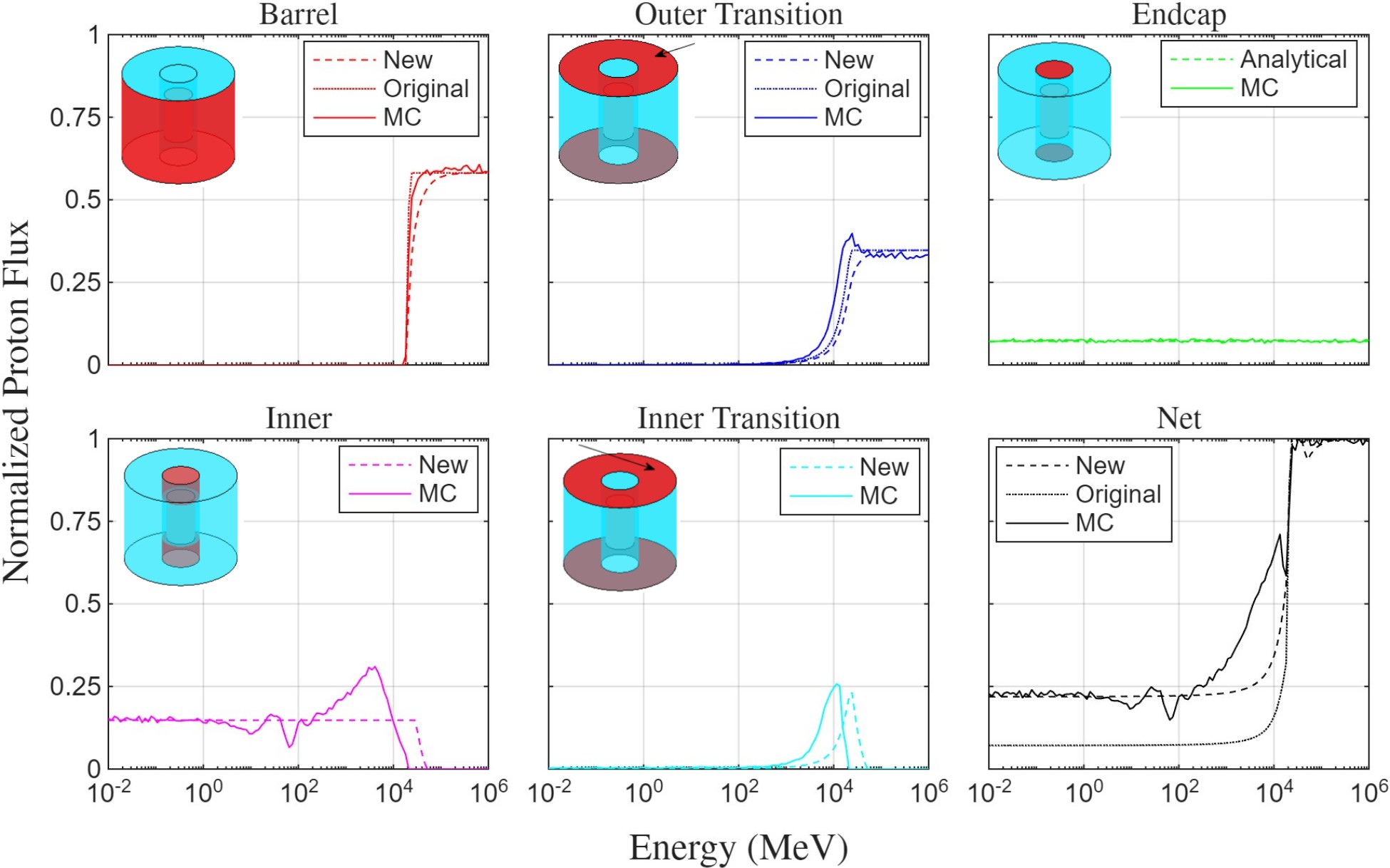}
    \caption{GCR proton flux through each of the shield surfaces calculated by the original analytical model, updated analytical model, and a Monte Carlo simulation for a 10 T $\times$ 10 m shield design, normalized by unattenuated flux at that energy. The original model does not account for flux through the inner and inner transition regions, and so underpredicts net flux at low energies.}\label{fig:shield_regions}
\end{figure}

There are four surfaces through which GCR particles can enter the shield: barrel, endcap, inner, and transition (Figure~\ref{fig:shield_regions}). Barrel particles initially strike the shield along its outermost wall $r=r_o$ and experience a high-pass filtering effect, where only particles with sufficiently high rigidities will penetrate through to the inside of the shield. Endcap particles enter through the very top or bottom of the shield with $r<r_i$ and do not ever encounter the magnetic field, and so maintain the same spectral characteristics as the unattenuated GCR spectrum. Similarly to endcap particles, inner particles enter through the ends of the shield, but instead intersect the inner wall of the shield ($r=r_i$) prior to reaching the habitat and can rebound off the shield back into the habitat. These particles experience a low-pass filtering effect: particles of sufficiently high rigidity will penetrate through to the outside wall of the shield, while low energy particles will be deflected back into the inside of the shield. Transition particles strike the shield along its top or bottom surface with $r_i < r < r_o$. Like the barrel surface, the transition surface acts as a high-pass filter, but only for particles whose velocities are directed towards the center of the shield at the moment of intersection (i.e., $\dot{r}<0$, hereafter referred to as ``outer transition'' particles). In the other case (``inner transition'' particles), where particle velocity is directed outward at the moment of intersection ($\dot{r}>0$), the transition surface instead acts as a band-pass filter: if particle energy is sufficiently high, it will penetrate through to the outside of the shield (much like the inner case), while particles with sufficiently low energy will penetrate neither the inner nor outer wall, and instead will remain within the field surface and move helically until they pass out of the top or bottom of the shield.

In the original model, the derivation of magnetic attenuation of incident GCR flux was greatly simplified by the conservative assumption that all particles approach along the most penetrating incidence angle $\alpha$. Furthermore, by integrating over the polar approach angle $\theta$ instead of the area elements of the shield, the original model only considered those particles approaching from directions that would have intersected the phantom \textit{had the shield not been present}; it does not account for those particles whose initial approaches would not have originally reached the phantom, but whose motion within the shield causes their deflection into the phantom. In other words, the original model only considered particle flux through the barrel, outer transition, and endcap surfaces, neglecting those particles passing through the inner transition surface and inner surface. 

In the updated model, the magnetic attenuation component is fully generalized. For each surface, an analytical expression for the GCR flux from an arbitrary infinitessimal area element on that surface onto a small spherical water phantom located within the habitat was derived. Surface integrals of these cutoff expressions over each surface are then numerically computed and used to construct the total attenuated GCR flux reaching the phantom through that surface. Detailed derivations of the surface flux integrals are available in the appendix.

\subsubsection{Magnetic Model Validation}

The new magnetic attenuation formulation was validated using Monte Carlo simulation. A simulation environment was constructed wherein GCR protons are transported through an annular cylindrical magnetic field of specifiable size and magnetic flux density. Protons ranging in energy from $10^{-2}$ to $10^{6}$ MeV are generated uniformly on the surface of a sphere of radius $r_{\text{inf}}=\text{100 m}$ surrounding the shield. The protons' velocities are initially directed inward towards the shield (i.e., antiparallel to the spherical $\hat{r}$ direction), then randomly rotated slightly away from the $\hat{r}$ to approximate an isotropic distribution. Particles are propagated using the Boris algorithm for pure magnetic fields \citep{boris_relativistic_1971} until they leave the vicinity of the shield or strike a small spherical phantom located at the origin. Additional details on the simulation construction are available in the appendix. A total of 250 million particles (2 million at each energy level) were simulated; phantom hits were tallied and normalized by the probability of a phantom hit for an uncharged particle under the same simulation conditions (Figure~\ref{fig:shield_regions}). Generally, the Monte Carlo results show better agreement with the new model across the range of energies simulated, particularly at lower energies. The new model still somewhat underpredicts phantom strikes at intermediate energies, but outperforms the original model in predicting the effect of magnetic attenuation on the incident GCR spectrum.

\subsection{Spatial Dose Variation}

The modified radiation environment within a shield varies as a function of position within the protected volume \citep{washburn_model_2013}. The original model only considered the dose along the axial centerline of the shield. Here, the rederived expressions for the GCR environment inside the habitat are fully generalized and capable of providing the approximate GCR spectrum at any point within the shielded habitat. This capability could be used in conjunction with a higher-fidelity human phantom to produce an estimate of tissue-specific doses and whole-body effective doses. It is noted that for points along the axial centerline, the $\phi$ dependence in the double integrals that describe the magnetically-attenuated GCR flux (Eqs.~(\ref{eqn:APP_FI_barrel}-\ref{eqn:APP_rlim})) is eliminated. This reduces them to single integrals over the radial or axial coordinate, reducing their computational cost by more than two orders of magnitude. Accordingly, the full spatial dose distribution should only be sparingly calculated. As a demonstration, the spatial distribution is calculated for a 20 m tall, \mbox{3 T $\times$ 5 m} shield, which is later shown to be the minimal-mass shield capable of reducing doses below 250 mSv/yr.

\section{Results}

\subsection{Field Uniformity Correction}
\begin{figure}
  \centering
   \includegraphics[width=0.5\textwidth]{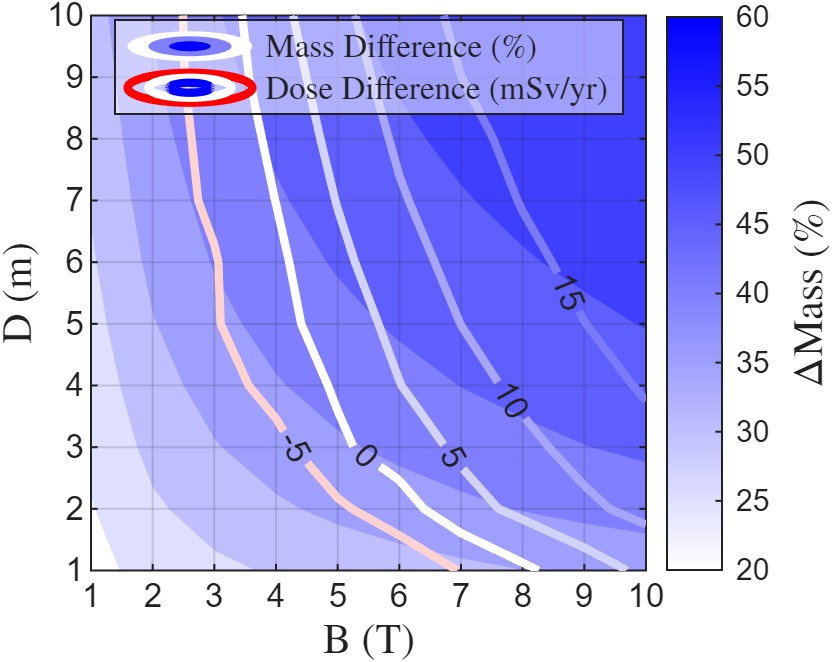}
    \caption{Effect of the field uniformity scaling factor on system mass (solid contours, in 5\% intervals) and dose equivalent (line contours, 5 mSv/yr intervals).}\label{fig:gap_scaling_result}
\end{figure}

Increasing the magnitude of the field produced by the coils ensures that the field averaging process does not artificially inflate the performance of the shield. The scaled flux density results in an additional system mass penalties. In turn, the extra mass serves as added passive shielding. This extra mass lowers the radiation dose until the effective thickness of the passive mass of the shield reaches a critical threshold of about 40 $\mathrm{g/cm}^2$, above which doses tend to increase with additional passive shielding (Figure~\ref{fig:gap_scaling_result}) due to neutron production within the material \citep{norman_early_2017}.

\subsection{Magnetic Attenuation Improvements}

\begin{figure}
  \centering
   \includegraphics[width=\textwidth]{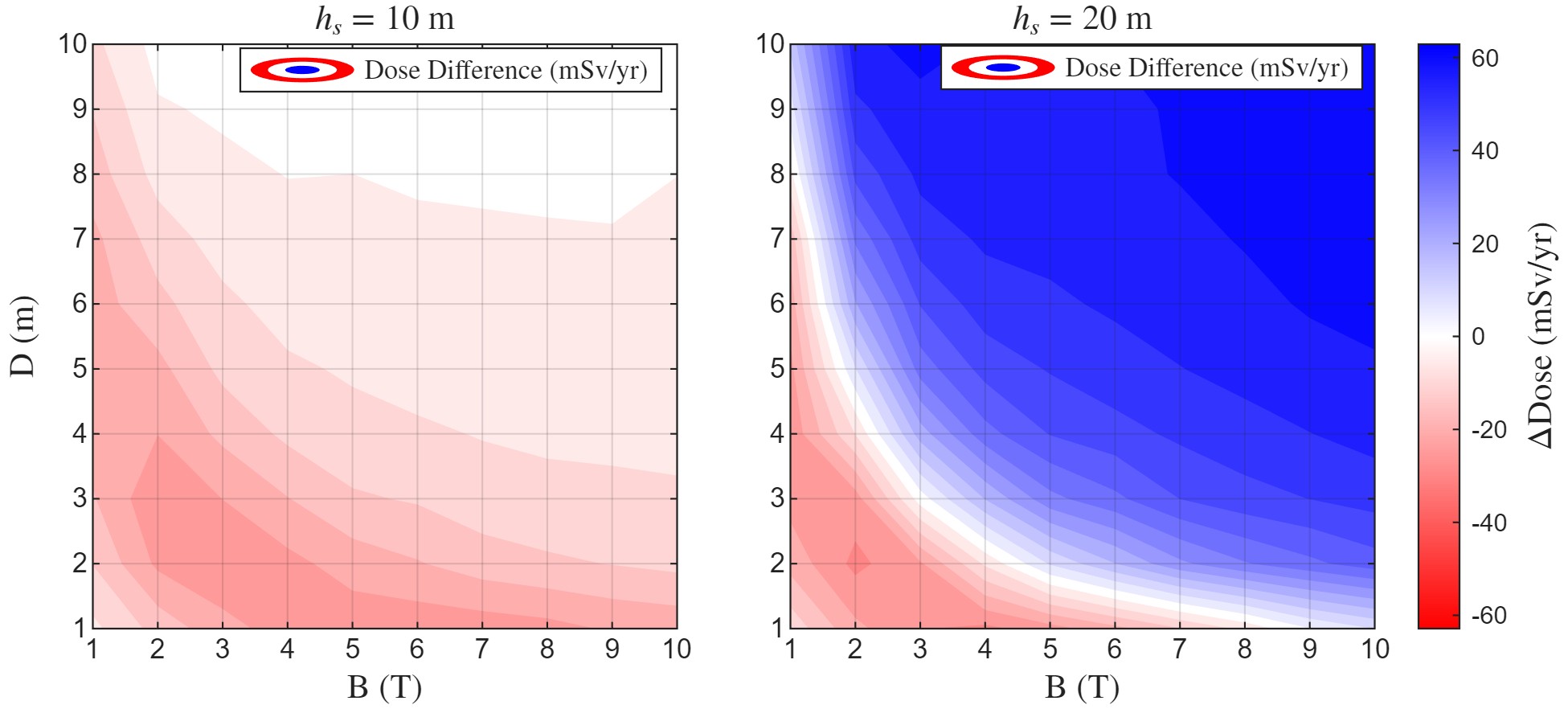}
    \caption{Change in predicted dose equivalent using updated magnetic model for 10 m (left) and 20 m tall (right) shields with 5 mSv/yr contours.}\label{fig:isotropy_result}
\end{figure}

The effects of the improvements to the magnetic attenuation computations are visualized for a range of shield designs (Figure~\ref{fig:isotropy_result}). The elimination of the original model's conservative assumption that all particles approach along the most penetrating incidence angle (i.e., $\alpha=0$) reduces estimated doses across the range of designs considered. However, the inclusion of particle flux through the inner and inner transition surfaces offsets this dose reduction. Increasing the size and flux density of the shield increases the size of the inner and inner transition surfaces and the cutoff energy below which particles are able to penetrate through these surfaces. Accordingly, the contribution to the total dose of inner and inner transition particles becomes increasingly relevant and ultimately outweighs the dose reduction associated with generalizing the incidence angle $\alpha$. For a 20 m tall, \mbox{3 T $\times$ 5 m} shield, the inner and inner transition surfaces respectively contribute 44.6 and 20.6 mSv/yr to the total dose of 246 mSv/yr. 

\subsection{Spatial Dose Variation}

\begin{figure}
  \centering
   \includegraphics[width=0.5\textwidth]{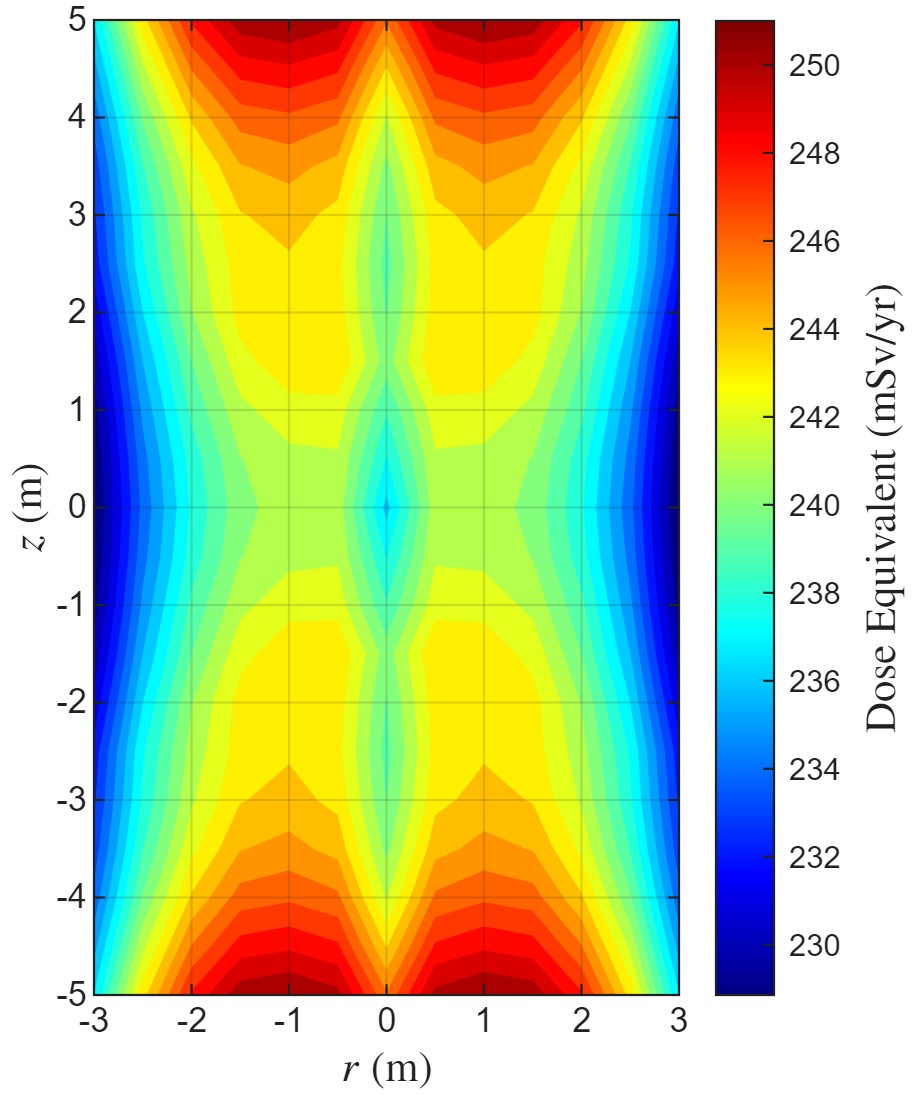}
    \caption{Spatial variation in solar minimum annual dose equivalent within a habitat protected by a 20 m tall, 3 T $\times$ 5 m solenoidal magnetic shield. The volume-averaged dose equivalent within the habitat is 240 mSv/yr. The dose equivalent varies by roughly 10\% within the volume of the habitat.}\label{fig:spatialdose}
\end{figure}

The spatial variation of dose within the habitat protected by a 20 m tall, 3 T $\times$ 5 m solenoidal shield is visualized (Figure~\ref{fig:spatialdose}). Annual dose equivalent at the standard test point within the shield (along the axial centerline at the end of the habitat) is 246 mSv/yr, while the volume-averaged dose equivalent throughout the habitat is 240 mSv/yr, a difference of approximately 3\%. While these differences are minor, other designs are likely to have more extreme spatial variations in dose equivalent, particularly for larger habitats and smaller shields. Future efforts to characterize spatial dose variations across a broader range of shield and habitat designs will be important to ensure that the standard axial test point used in this article remains an appropriate, representative choice for characterizing shield performance.

\subsection{Net Results}
\begin{figure}
  \centering
   \includegraphics[width=\textwidth]{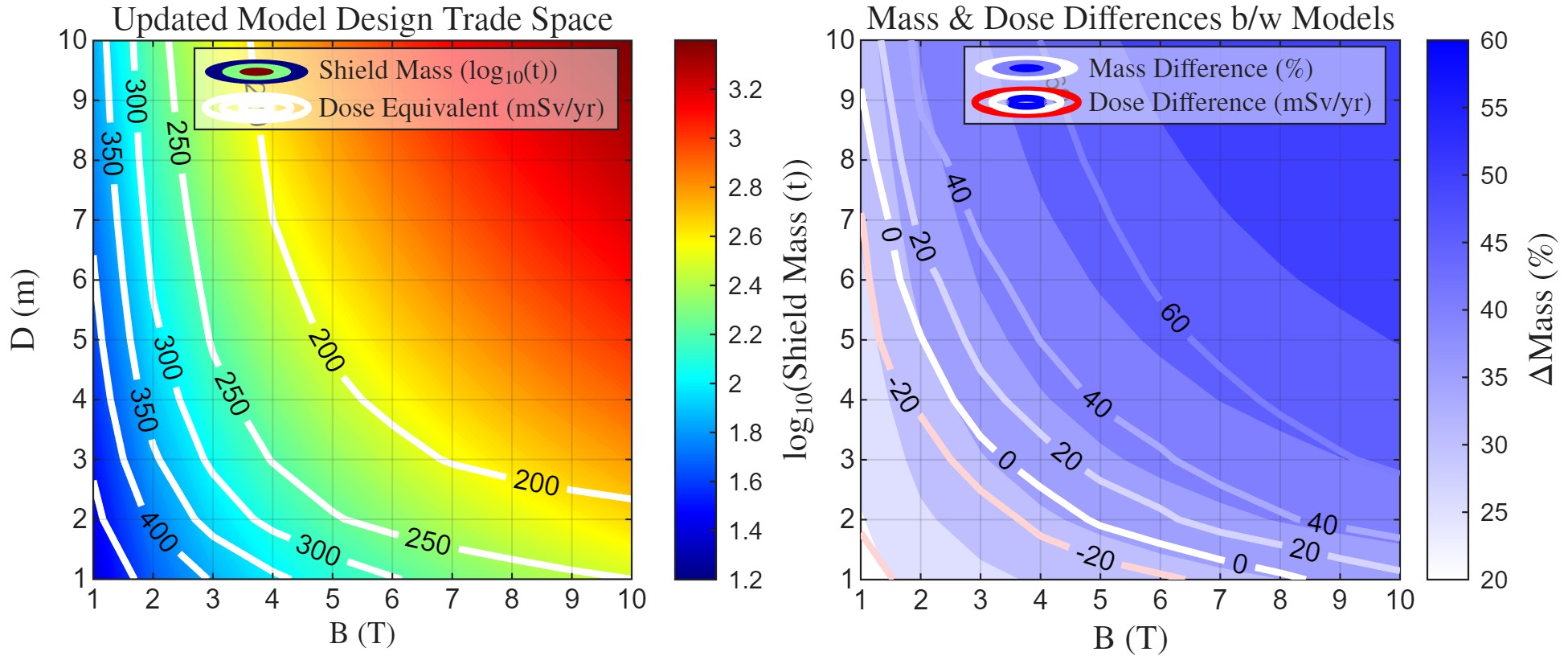}
    \caption{Mass and annual dose equivalent contours for 20 m tall shields with varied shield thickness and flux density as predicted by the updated model (left). Differences in predicted mass and annual dose equivalent between the original and updated models (right).}\label{fig:combined}
\end{figure}

Estimated system masses and annual dose equivalents under the updated model with all aforementioned changes (Figure~\ref{fig:combined}, left) and its difference compared to the original model (right) are visualized for 20 m tall shields. The combined effects of the updates to the model result in reductions in estimated dose equivalents for shields with bending powers below about 9 T-m, while stronger shields have higher predicted doses. The dose reductions for weak shields are driven by both the removal of the assumption that all particles approach along the most penetrating horizontal incidence angle $\alpha$ and the increased passive shielding associated with the annular cylinder geometric scaling factor. The dose increases for strong shields are driven by the inclusion of low energy GCR particles entering the shield through the inner and inner transition surfaces. For example, for a 20 m tall, 1 T $\times$ 1 m shield achieving a total dose equivalent of 477 mSv/yr, less than 0.1\% of the total dose equivalent is contributed by inner and inner transition surface particles, while for a 20 m tall, 10 T $\times$ 10 m shield, that percentage jumps to nearly 40\%.

It is apparent that the diminishing returns of increasing shield bending power beyond about 20 T-m are much more significant than predicted by the original model. Whereas the original model predicted an annual dose equivalent of approximately 80 mSv/yr for a 20 m tall, 10 T $\times$ 10 m shield (approximately an 84\% reduction relative to the dose at that point protected only by the habitat's pressure vessel), the new model predicts a dose equivalent of 188 mSv/yr, a reduction of only 63\%. This finding generally agrees with a prior analysis of an infinitely-tall toroidal magnetic shield \citep{musenich_limits_2018}, where it was argued that magnetic shields are not likely to be viable as a sole countermeasure if GCR dose reductions on the order of 90\% relative to free space are desired. However, smaller shields of about 10\textendash20 T-m still provide a significantly greater GCR dose reduction than equivalent-mass passive shields, and so active magnetic shields are still worth investigating for inclusion in a larger suite of radiation countermeasures.

\begin{figure}[h]
  \centering
   \includegraphics[width=\textwidth]{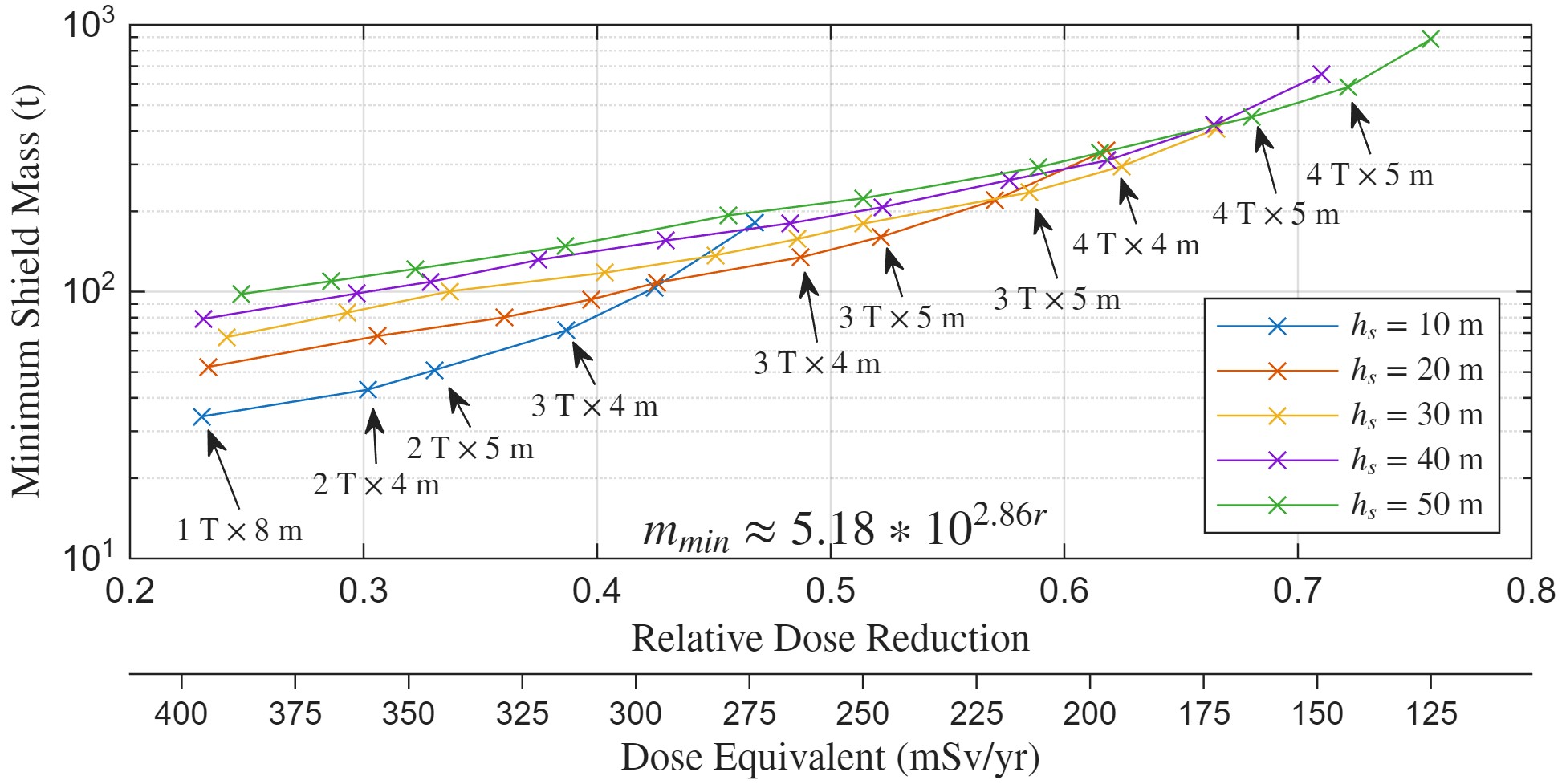}
    \caption{Minimum mass shields of various heights that achieve specified dose equivalents. Relative dose reduction is defined with respect to the GCR dose received at a point protected only by the habitat pressure vessel (514 mSv/yr).}\label{fig:massopt}
\end{figure}

An optimal shield achieves a specified reduction in dose equivalent at a minimum mass. Identified optimal designs for shields ranging in height from 10\textendash50 m and bending power from 1\textendash100 T-m are visualized (Figure~\ref{fig:massopt}). Across the range of dose reductions considered, larger shields with weaker magnetic fields are more mass efficient than smaller shields with stronger magnetic fields, consistent with the original model. This data also reveals an apparent lower bound on shield mass with respect to dose reduction. This exponential relationship could serve as a useful benchmark for broadly comparing the performance of solenoidal shields to other archetypal shield configurations without necessitating the selection of specific design variables. It is envisioned that ongoing work to produce a similar performance evaluation model for toroidal and hybrid shields will ultimately yield a similar relationship that can be used to characterize the relative performance of these broad classes of shield designs across an entire trade space.

\begin{figure}
  \centering
   \includegraphics[width=\textwidth]{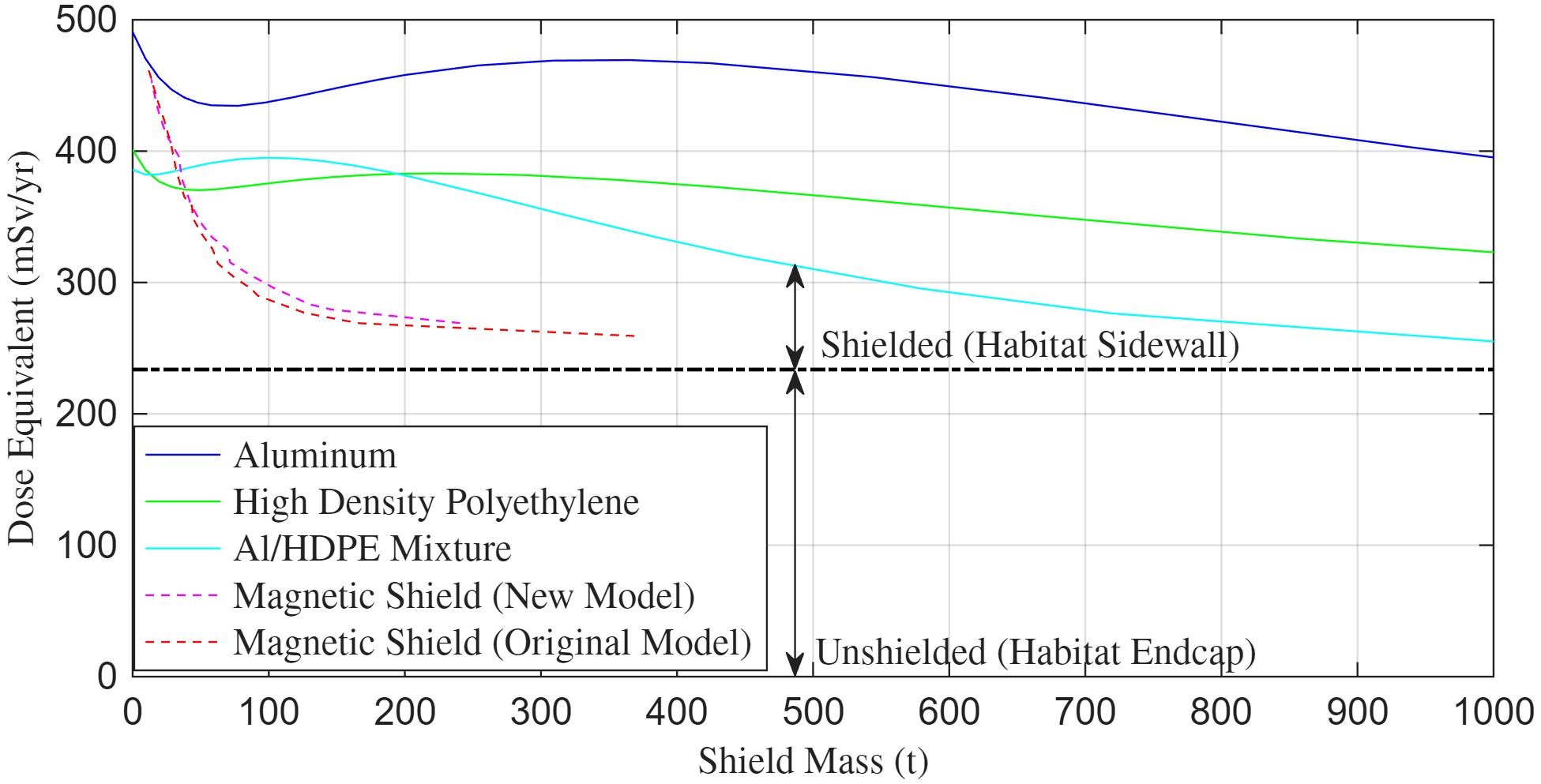}
    \caption{Performance comparison of solenoidal magnetic shields to passive shields consisting of aluminum, high density polyethylene, and an equal mixture of both materials. The protected habitat has the same characteristics as described in Table~\ref{tab:defaults}. The shields considered here are all 10 m tall and protect only the sidewall of the habitat, to facilitate comparison by ensuring that unshielded endcap contributions are the same for all designs considered.}\label{fig:passiveactive}
\end{figure}

While the updated model generally predicts worse mass-normalized performance of solenoidal shields than the original model, these shields are still shown to be significantly more performant than mass-equivalent passive shields (Figure~\ref{fig:passiveactive}). For the representative habitat described in Table~\ref{tab:defaults}, a supplementary active shield universally outperforms aluminum passive shields across all dose thresholds, and outperforms passive shields utilizing polyethylene when the requisite annual dose equivalent is below about 350 mSv/yr (approximately a 60\% reduction relative to the free space dose).

\section{Conclusions}

The improved rapid analysis model presented here enables a more accurate quantification of the performance of solenoidal radiation shields without appreciably increasing the computational cost of the model. Introducing the gap scaling factor corrects a significant positive shield performance bias present in the original model. Improvements to the semi-analytical magnetic GCR attenuation model enable a more accurate description of the attenuated GCR environment reaching the habiat and have been verified using a Monte Carlo simulation of charged particle motion within a shield. Finally, the fully generalized magnetic flux integrals are capable of producing estimates of dose equivalent at any location within the habitat, not just along the axial centerline, which enables more thorough evaluation of shield performance. The combined effects of these updates to the rapid analysis model suggest that the original model underestimated the performance of small, weak shields and significantly overestimated the performance of larger shields with bending powers above 10 T-m. The updated model presented here provides better agreement with existing literature regarding the performance of magnetic shields of varying bending power, and suggest that significant reductions in astronaut GCR doses relative to free space by an order of magnitude or more are unlikely to be achieved using magnetic shields alone. 

Ongoing efforts to further improve the utility of the model include an incorporation of a number of technological advancements that have been realized since the original model was developed in 2014, which will provide an updated understanding of present magnetic shielding capabilities using state-of-the-art technologies. Future planned efforts include further validation of the accuracy of the model via high-fidelity Monte Carlo simulation of selected shield designs, identification of other key performance-driving design variables, and an extension of the modeling approach to toroidal and hybrid shield designs, whose performance may vary significantly from the solenoidal designs considered here. 


\section*{Acknowledgments}
The authors thank Dr.\ Scott Washburn for helping to clarify various aspects of the shield mass estimation process.
This material is based upon work supported by the National Science Foundation Graduate Research Fellowship Program under Grant No. DGE 2040434. Any opinions, findings, and conclusions or recommendations expressed in this material are those of the author and do not necessarily reflect the views of the National Science Foundation.

\clearpage
\section*{Appendix}
\appendix


\section{Example of Mass Estimation Process}
\renewcommand{\theequation}{A.\arabic{equation}}
\renewcommand{\thefigure}{A\arabic{figure}} 
\renewcommand{\thetable}{A\arabic{table}} 
\setcounter{equation}{0}
\setcounter{figure}{0}

Here we step through an example derivation of system mass as a function of shield design variables and technology performance parameters. This example is reproduced with minor adjustments from \citet{washburn_model_2013}. Consider the radial support component of the shield's structure subsystem. The magnetic pressure on an ideal, long, thin-walled cylindrical coil is given by

\begin{equation}
  P_B = \frac{B^2}{2\mu_0}
  \label{eqn:magnetic_pressure}
\end{equation}

\citep{iwasa_case_1995}. Pressure inside a cylinder is related to hoop stress by 

\begin{equation}
  \sigma_t = \frac{P_B D}{2 t_w}
  \label{eqn:hoop_stress}
\end{equation}

where $t_w$ is the thickness of the coil. Substituting Eq.\ (\ref{eqn:magnetic_pressure}) into Eq.\ (\ref{eqn:hoop_stress}) and including a factor of safety $F_s$ yields an equation for the maximum allowable hoop stress

\begin{equation}
  \sigma_{t,\max} = \frac{B^2 D F_S}{4 \mu_0 t_w}
  \label{eqn:magnetic_pressure_hoop_stress}
\end{equation}

Eq.\ (\ref{eqn:magnetic_pressure_hoop_stress}) can be rearranged for the wall thickness $t_w$, from which an expression for the volume and mass of the radial support material necessary for each coil can be obtained

\begin{equation}
  m_{\text{radial}} = \rho_{\text{radial}}V_{\text{radial}} = \frac{\rho_{\text{radial}} B^2 D^2 \pi h_s F_S}{4 \mu_0 \sigma_{u,\text{radial}}}
  \label{eqn:radial_structure_mass}
\end{equation}

where $\rho_{\text{radial}}$ and $\sigma_{u,\text{radial}}$ are the mass density and the ultimate tensile strength of the radial support material. Eq.~(\ref{eqn:radial_structure_mass}) provides a simple relationship between subsystem mass, key shield design parameters (magnetic flux density $B$, depth $D$ and height $h_s$) and technology performance parameters (material tensile strength $\sigma_u$ and mass density $\rho$), which enables evaluation of the influence of shield design and technology advancement on overall shield mass. When combined with similar relationships derived for other key subsystem components, subsytem-level mass estimates are produced, which in turn are summed to produce an overall mass estimate for the shield.

\section{Derivation of magnetic attenuation}\label{}

\renewcommand{\theequation}{B.\arabic{equation}}
\renewcommand{\thefigure}{B\arabic{figure}} 
\renewcommand{\thetable}{B\arabic{table}} 
\setcounter{equation}{0}
\setcounter{figure}{0}

One of the primary shortcomings of the original model was its inability to accurately predict the GCR environment inside a protected habitat. Specifically, the original model first assumes that all approaching particles are incident upon the shield along the most penetrating horizontal incidence angle $\alpha=0$. Furthermore, the model is unable to account for inner and inner transition surface particles, since the double integrals describing the particle flux in the original model are constructed using spherical approach angles $\theta$ and $\phi$, which are defined with respect to the phantom and therefore fail to account for particles with $\dot{r}>0$. While these approximations do greatly simplify the derivation of the magnetic cutoff relationships and their numerical complexity, they result in substantially different findings regarding the ideal configuration of a solenoidal shield. By reframing the flux integrals as surface integrals over the various surfaces of the shield, the magnetic attenuation to the incident GCR spectrum can be more accurately described. Here, the derivation of the updated surface flux integrals is presented.

\subsection{Barrel Surface}

Consider the scenario visualized in Figure~\ref{fig:shield_layout}. An annular cylindrical field of height $h_s$, inner radius $r_i$, outer radius $r_o$, and uniform flux density $B$ throughout is struck along the barrel surface by a GCR particle of rest mass $m$ and charge $q$ at polar angle $\theta$ defined with respect to the $+Z$ axis and horizontal incidence angle $\alpha$ defined with respect to the line tangent to $r_o$ at the point of intersection. That particle travels along an arc of radius $r_g$ until it reaches $r_i$, at which point it travels in a straight line until intersecting a spherical phantom located at radius $r_p$ along an azimuth $\phi$ defined with respect to $r_p$. From the right graphic of Figure~\ref{fig:shield_layout}, a purely geometrical relationship between $r_g$ and the shield dimensions and approach conditions can be computed using the law of cosines 

\begin{equation}
  r_g = \frac{r_o^2-r_i^2}{2(r_o\cos(\alpha) + r_p\cos(\phi))}
  \label{eqn:APP_rg_geometrical}
\end{equation}

Now, assuming that the phantom at $r_p$ has some small, finite radius $R$, a range of incidence angles $(\alpha-\Delta\alpha/2,\alpha+\Delta\alpha/2)$ and $(\theta-\Delta\theta/2,\theta+\Delta\theta/2)$ can yield a trajectory that ultimately strikes the phantom. The small angle approximation can be used to update Eq.\ (\ref{eqn:APP_rg_geometrical}) to 

\begin{equation}
  r_g = \frac{r_o^2-r_i^2}{2(r_o\cos(\alpha) + (r_p\cos(\phi)\pm R))}
  \label{eqn:APP_rg_geometrical_R_barrel}
\end{equation}

which in turn can be solved for the window of horizontal incidence angles which yield trajectories that ultimately intersect the phantom 

\begin{equation}
  \Delta \alpha = \frac{2R}{r_o}
    \label{eqn:APP_delta_alpha_barrel}
\end{equation}

The polar angle window can be computed as 

\begin{equation}
  \Delta\theta = \frac{2R}{\sqrt{\mathrm{r_o'}(r_p,\phi)^2 + (z-h_p)^2}}
  \label{eqn:APP_delta_theta_barrel}
\end{equation}

where $z$ is the height at which the particle initially strikes the shield barrel region and where $\mathrm{r'}(r_p,\phi)$ is the distance to the edge of a circle of radius $r$ from a point located a distance $r_p$ from the center of $r$ along an azimuth $\phi$, i.e.,

\begin{equation}
  \mathrm{r'}(r_p,\phi) = -r_p\cos(\phi) + \sqrt{r^2 - (r_p\sin(\phi))^2}
  \label{eqn:APP_rp}
\end{equation}

Together, these expressions for $\Delta\alpha$ and $\Delta\theta$ can used to compute the total solid angle from which particles incident upon an infinitessimal area element $dA = r_o d\phi dz$ located at $(r_o,z)$ can approach and strike the phantom 

\begin{align}
  \Delta\Omega &= \frac{\pi}{4}\int_{\alpha-\Delta\alpha/2}^{\alpha+\Delta\alpha/2} \int_{\theta-\Delta\theta/2}^{\theta+\Delta\theta/2} \sin(\theta')\mathrm{d}\theta'\mathrm{d}\alpha' \label{eqn:APP_delta_Omega_barrel} \\
  &\approx \frac{\pi}{4}\Delta\alpha \sin(\theta)\Delta\theta \notag \\
  &=\frac{\pi R}{2r_o} \left( \frac{\mathrm{r_o'}(r_p,\phi)}{\sqrt{\mathrm{r_o'}(r_p,\phi)^2 + (z-h_p)^2}}  \right) \left( \frac{2R}{\sqrt{\mathrm{r_o'}(r_p,\phi)^2 + (z-h_p)^2}} \right) \notag \\
  &= \frac{\pi R^2 \mathrm{r_o'}(r_p,\phi)}{r_o\left(\mathrm{r_o'}(r_p,\phi)^2 + (z-h_p)^2\right)} \notag
\end{align}

where the factor of $\pi/4$ converts from a rectangular solid angle element to an elliptical element. The cylindrical area element $\mathrm{d}A$ must be scaled by a factor describing the projection of $\mathrm{d}A$ with respect to the direction of the solid angle element 

\begin{align}
  \mathrm{d}A_{\text{proj}} &= \sin(\theta)\sin(\alpha)dA  \label{eqn:APP_dA_proj_barrel} \\
  &= \frac{\mathrm{r_o'}(r_p,\phi) r_o }{\sqrt{\mathrm{r_o'}(r_p,\phi)^2 + (z-h_p)^2}}\mathrm{Re}\left[\sqrt{1- \left( \frac{r_o^2 - r_i^2 -2r_p\cos(\phi) r_g}{2 r_o r_g}\right)^2}\right]\mathrm{d}\phi\mathrm{d}z \notag
\end{align}

where the $\mathrm{Re}[\cdots]$ term encodes the energy cutoff conditions by reexpressing $\sin(\alpha)$ as $\sqrt{1-\cos(\alpha)^2}$. The cutoff energy occurs below the point where an approaching particle cannot breach the shield even with $\alpha=0$. Putting it all together and selecting $R$ such that the phantom has a surface area of 1 $\text{cm}^2$, the surface integral describing all GCR flux reaching point $(r=r_p,\phi=0,z=h_p)$ through the barrel surface is given by

\begin{align}
  \Phi_{\text{b}}(K) &= \Phi(K) \int_{0}^{2\pi}\int_{-\frac{h_s}{2}}^{\frac{h_s}{2}} \frac{\mathrm{d}\Omega}{4\pi} \mathrm{d}A_{\text{proj}} \label{eqn:APP_FI_barrel} \\
  &= \Phi(K) \int_{0}^{2\pi}\int_{-\frac{h_s}{2}}^{\frac{h_s}{2}}  \frac{{\mathrm{r_o'}(r_p,\phi)}^2}{\left({{\mathrm{r_o'}(r_p,\phi)}}^2 + \left(z-h_p\right)^2\right)^{3/2}} \notag \\
  &\hspace{1cm} \mathrm{Re}\left[\sqrt{1 - \left(\frac{{r_o}^2 - {r_i}^2 - 2r_p\cos(\phi) \mathrm{r_g}(K,r_o,\phi,z)}{2 r_o \mathrm{r_g}(K,r_o,\phi,z)}\right)^2}\right] \mathrm{d}z \mathrm{d}\phi \notag
\end{align}

where $\Phi(K)$ is the differential omnidirectional GCR flux of a particle of energy $K$ in $\text{cm}^{-2} \text{day}^{-1} \text{MeV}^{-1}$ and $r_g$ is the gyroradius of a particle entering the shield at a point $(r,\phi,z)$ on its surface with an initial velocity directed towards point $(r_p,0,h_p$) 

\begin{equation}
  \mathrm{r_g}(K,r,\phi,z) = \frac{m_0 c }{qB}\sqrt{\left(1+\frac{K}{m_0c^2}\right)^{2} - 1} \left(\frac{\mathrm{r'}(r_p,\phi)}{\sqrt{\mathrm{r'}(r_p,\phi)^2 + \left(z - h_p\right)^2}}\right)
  \label{eqn:APP_E_to_rg}
\end{equation}

In the limiting case where $r_p=0$ (i.e., the phantom is located on the central axis of the habitat), $\mathrm{r'}(r_p,\phi)=r$ and all $\phi$ dependence is removed from the integrand, reducing Eq.\ (\ref{eqn:APP_FI_barrel}) and all the following flux integrals to a single integral over the radial or axial coordinate and enabling a significant reduction in computational complexity.

The derivation of the remaining surface integrals follow a similar approach wherein an expression for $r_g$ is derived geometrically and used to form an expression for $\Delta\alpha$ and $\mathrm{d}\Omega$. The projection factor $\mathrm{d}A_{\text{proj}}$ is then computed, then the final surface integral is constructed as in to the first step of Eq.\ (\ref{eqn:APP_FI_barrel}).

\subsection{Outer Transition}

The derivation of GCR flux through the outer transition surface closely mirrors the barrel surface. The geometrical expression for the gyroradius is equivalent to Eq.\ (\ref{eqn:APP_rg_geometrical_R_barrel}) but with $r_o \rightarrow r$

\begin{equation}
  r_g = \frac{r^2-r_i^2}{2(r\cos(\alpha) + (r_p\cos(\phi)\pm R))}
  \label{eqn:APP_rg_geometrical_R_outtrans}
\end{equation}

which yields 

\begin{equation}
  \Delta \alpha = \frac{2R}{r}
    \label{eqn:APP_delta_alpha_trans}
\end{equation}

where $r$ is the radial position of the cylindrical surface element $\mathrm{d}A$. The polar angle window is given by

\begin{equation}
  \Delta\theta = \frac{4R}{\sqrt{4r^2 + (h_s \pm 2h_p)^2}}
  \label{eqn:APP_delta_theta_outtrans}
\end{equation}

where the positive case applies to the bottom surface of the shield, and the negative to the top. The solid angle window is therefore given by 

\begin{align}
  \Delta\Omega &= \frac{\pi}{4}\int_{\alpha-\Delta\alpha/2}^{\alpha+\Delta\alpha/2} \int_{\theta-\Delta\theta/2}^{\theta+\Delta\theta/2} \sin(\theta')\mathrm{d}\theta'\mathrm{d}\alpha' \label{eqn:APP_delta_Omega_outtrans} \\
  &\approx \frac{\pi}{4}\Delta\alpha \sin(\theta)\Delta\theta \notag \\
  &=\frac{\pi R}{2r} \left( \frac{2r}{\sqrt{4r^2 + (h_s \pm 2h_p)^2}}  \right) \left( \frac{4R}{\sqrt{4r^2 + (h_s \pm 2h_p)^2}} \right) \notag \\
  &= \frac{4\pi R^2}{\left(4r^2 + (h_s \pm 2h_p)^2\right)} \notag
\end{align}

The projected area element is given by 
\begin{align}
  \mathrm{d}A_{\text{proj}} &= \cos(\theta)\sin(\alpha)\mathrm{d}A \label{eqn:APP_dA_proj_outtrans} \\
  &= \frac{(h_s \pm 2h_p)}{\sqrt{4r^2 + (h_s \pm 2h_p)^2}} \notag\\
  &\hspace{1cm} \mathrm{Re}\left[ \sqrt{1 - \left( \frac{r^2 + r_p^2 - r_i^2 - 2 (r+\mathrm{r_g}(K,r,\phi,\mp\frac{h_s}{2})) r_p \cos(\phi)}{2 r \, \mathrm{r_g}(K,r,\phi,\mp\frac{h_s}{2})} \right)^2 }\right] r \mathrm{d}\phi \mathrm{d}z \notag
\end{align}

As before, choosing $R$ such that the spherical phantom has a surface area of 1 $\text{cm}^2$, the flux of a GCR species with energy $K$ reaching the phantom through the outer transition surface is given by 
\begin{samepage}
\begin{align}
  \Phi_{\text{t,o}}(K) &= \int_{0}^{2\pi}\int_{\mathrm{r_i'}(r_p,\phi)}^{\mathrm{r_o'}(r_p,\phi)} \frac{4 r \left(h_s \pm 2h_p\right) }{\left(4r^2 + \left(h_s \pm 2h_p\right)^2\right)^{3/2}}   \label{eqn:APP_FI_trans_outer}\\ 
  & \mathrm{Re} \left[\sqrt{1 - \left(\frac{ r^2 + r_p^2 - {r_i}^2 - 2(r+\mathrm{r_g}(K,r,\phi,\mp \frac{h_s}{2}))r_p\cos(\phi)}{2 r \,\mathrm{r_g}(K,r,\phi, \mp \frac{h_s}{2})}\right)^2}\right] \mathrm{d}r\mathrm{d}\phi  \notag
\end{align}
\end{samepage}

\subsection{Inner Surface}

\begin{figure}[h!]
  \centering
   \includegraphics[scale=0.5]{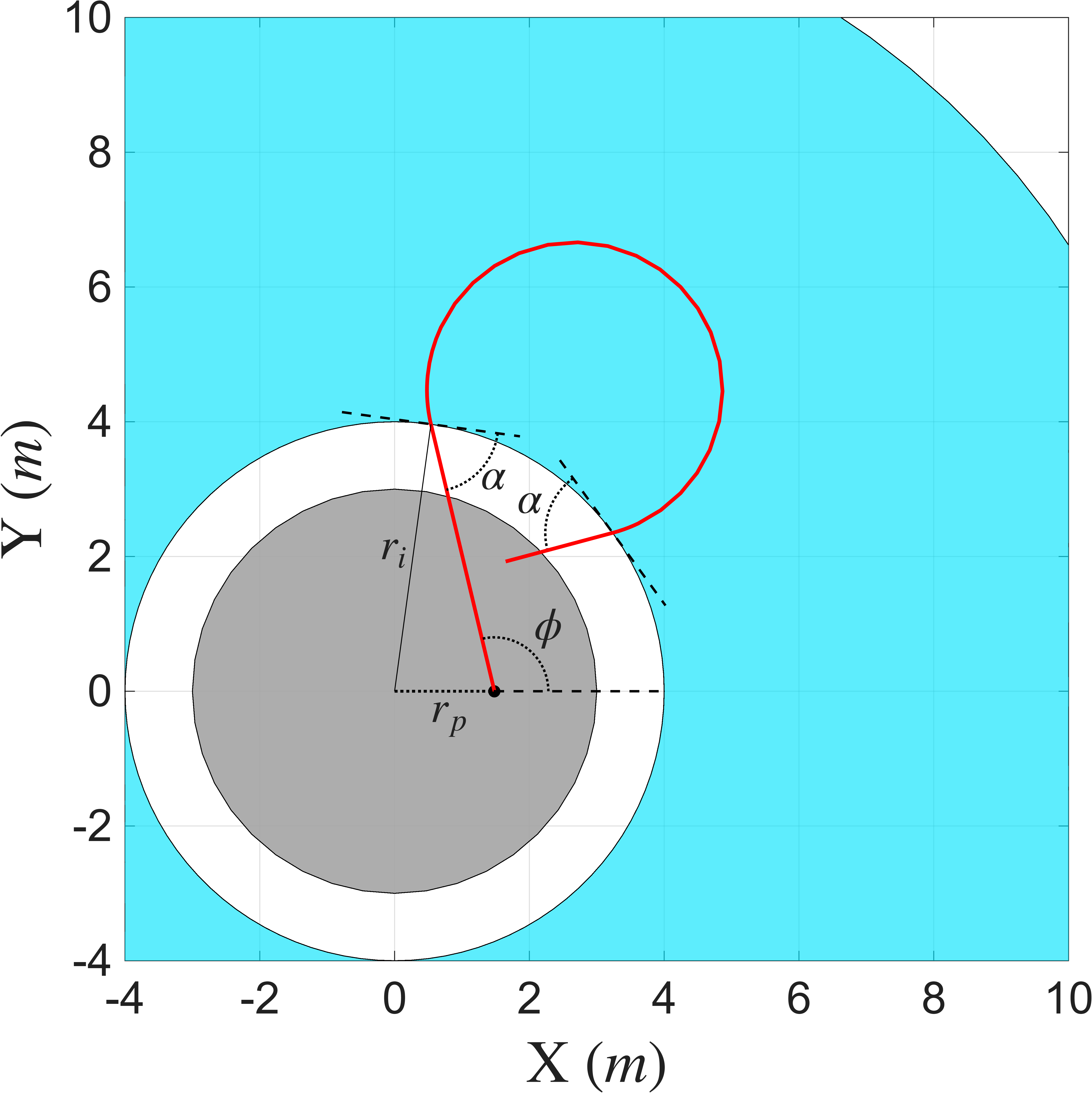}
    \caption{Motion of a particle entering the shield along the inner surface and striking the phantom.}\label{fig:APP_inner_rl}
\end{figure}

Particles entering the shield through the inner and inner transition surfaces behave differently from barrel and outer transition particles and thus the derivation proceeeds a bit differently. Consider a particle entering the shield through the inner surface (Figure~\ref{fig:APP_inner_rl})

The law of sines yields a simple expression for $\alpha$

\begin{equation}
  \cos(\alpha) = \frac{r_p }{r_i}\sin(\phi)
  \label{eqn:APP_alpha_inner}
\end{equation}

As before, expanding the phantom from a point to a small sphere of radius $R$ enables reexpressing Eq. (\ref{eqn:APP_alpha_inner}), which ultimately yields expressions for $\Delta\alpha$ and $\Delta\theta$

\begin{equation}
  \cos(\alpha) = \frac{r_p \pm R\csc(\phi)}{r_i}\sin(\phi)
  \label{eqn:APP_alpha_inner_R}
\end{equation}

which yields

\begin{equation}
  \Delta\alpha = \frac{2R}{r_i}
  \label{eqn:APP_delta_alpha_inner}
\end{equation}

and

\begin{equation}
  \Delta\theta = \frac{2R}{\sqrt{\mathrm{r_o'}(r_p,\phi)^2 + (z-h_p)^2}}
  \label{eqn:APP_delta_theta_inner}
\end{equation}

As before, $\Delta\alpha$ and $\Delta\theta$ can be used to compute the solid angle element $\Delta\Omega$

\begin{samepage}
\begin{align}
  \Delta\Omega &= \frac{\pi}{4}\int_{\alpha-\Delta\alpha/2}^{\alpha+\Delta\alpha/2} \int_{\theta-\Delta\theta/2}^{\theta+\Delta\theta/2} \sin(\theta')\mathrm{d}\theta'\mathrm{d}\alpha' \label{eqn:APP_delta_Omega_inner}\\
  &= \frac{\pi R^2 \mathrm{r_o'}(r_p,\phi)}{r_i\left(\mathrm{r_o'}(r_p,\phi)^2 + (z-h_p)^2\right)} \notag
\end{align}
\end{samepage}

The projected area element for the inner surface is given by

\begin{align}
  \mathrm{d}A_{\text{proj}} &= \sin(\theta)\sin(\alpha)\mathrm{d}A \label{eqn:APP_dA_proj_inner}\\
  &= \frac{\mathrm{r_o'}(r_p,\phi)}{\sqrt{\mathrm{r_o'}(r_p,\phi)^2 + (z-h_p)^2}} \sqrt{1-\left(\frac{r_p}{r_i}\sin(\phi)\right)^2} \mathrm{d}\phi \mathrm{d}z \notag
\end{align}

Eqs.~(\ref{eqn:APP_delta_Omega_inner}) and (\ref{eqn:APP_dA_proj_inner}) form the basis of the flux integral describing GCR flux through the inner surface 

\begin{samepage}
\begin{align}
  \Phi_{\text{i}}(K) &= \int_{0}^{2\pi}\int_{\mathrm{z}_{\text{min}}(K,\phi)}^{\frac{h_s}{2}} \frac{{\mathrm{r_o'}(r_p,\phi)}^2 } {\left({{\mathrm{r_o'}(r_p,\phi)}}^2 + \left(z-h_p\right)^2\right)^{3/2}}  \sqrt{1 - \left(\frac{r_p}{r_i}\sin(\phi)\right)^2} \mathrm{d}z \mathrm{d}\phi +  \label{eqn:APP_FI_inner}\\ 
  & \int_{0}^{2\pi}\int_{\frac{-h_s}{2}}^{\mathrm{z}_{\text{max}}(K,\phi)} \frac{{\mathrm{r_o'}(r_p,\phi)}^2 } {\left({{\mathrm{r_o'}(r_p,\phi)}}^2 + \left(z-h_p\right)^2\right)^{3/2}}  \sqrt{1 - \left(\frac{r_p}{r_i}\sin(\phi)\right)^2} \mathrm{d}z \mathrm{d}\phi \notag
\end{align}
\end{samepage}

where the $z$ integration limits are determined by either the particle's energy cutoff or shield self-shadowing effects

\begin{equation}
  \mathrm{z}_{\text{lim}}(K,\phi) = \text{clamp}\left(\max\left(z_{\text{lim,K}}(K,\phi), z_{\text{lim,s}}(\phi) \right),\pm\frac{h_h}{2}, \pm\frac{h_s}{2}   \right)
  \label{eqn:APP_zlim}
\end{equation}

where the $+$ case applies for $z_{\text{min}}$ and the $-$ case for $z_{\text{max}}$. The $\text{clamp}(x,a,b)$ restricts $x$ to lie in the interval $[a,b]$. The energy and shadowing cutoffs are respectively given by 

\begin{equation}
  \mathrm{z_{\text{lim},K}}(K,\phi) = h_p \pm \mathrm{Re}\left[\sqrt{ \left(\frac{2 \mathrm{r_g}(K,r_o,\phi,\pm h_p)\mathrm{r_o}(r_p,\phi)(r_o+r_p\sin(\phi))}{r_o^2-r_i^2} \right)^2 - \mathrm{r_o}(r_p,\phi)^2}\right]
  \label{eqn:APP_zlim_K}
\end{equation}

and

\begin{equation}
  \mathrm{z_{\text{lim},s}}(\phi) = \frac{ 2h_p(r_o+r_i) \pm h_s \mathrm{r_o'}(r_p,\phi)}{2(\mathrm{r_o'}(r_p,\phi)+r_o+r_i)}
  \label{eqn:APP_zlim_s}
\end{equation}

where, as before, the $+$ case applies for $z_{\text{min}}$ and the $-$ case for $z_{\text{max}}$. 

\subsection{Inner Transition}

\begin{figure}[h!]
  \centering
   \includegraphics[scale=0.5]{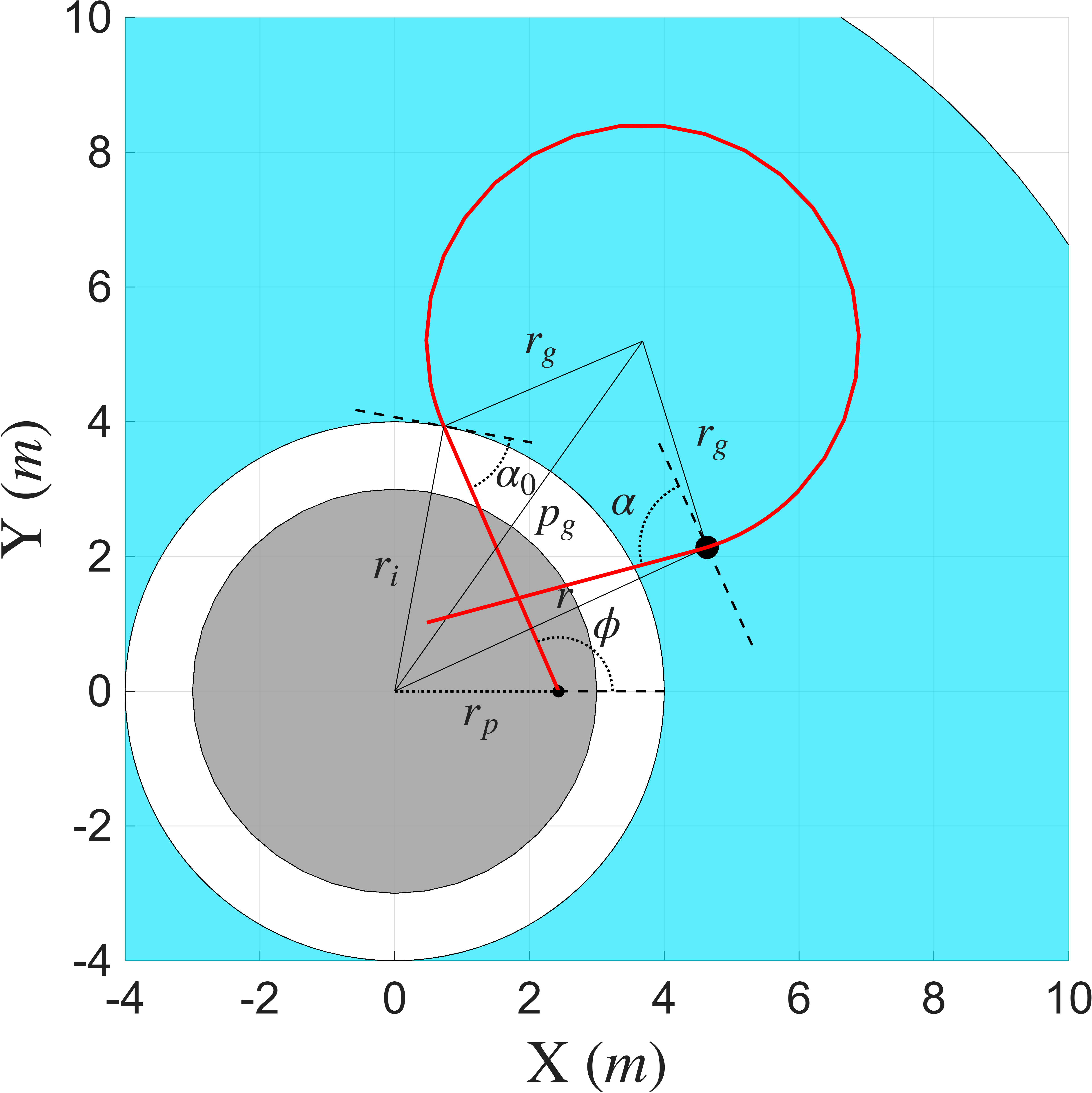}
    \caption{Motion of a particle entering the shield through the inner transition region at horizontal incidence angle $\alpha$ at a radius $r$ (black dot), moving through the shield and striking the phantom.}\label{fig:APP_innertrans_rl}
\end{figure}

The geometry of particle motion through the inner transition surface is the most complicated to analyze. Consider a particle entering the shield along the inner transition surface (Figure~\ref{fig:APP_innertrans_rl}). The law of cosines applied to the triangles $\Delta r_i r_g p_g$ and $\Delta r r_g p_g$ yields the following expression for $\alpha_0$

\begin{equation}
  \cos(\alpha_0) = \frac{\left(r^2 - r_i^2 \right) + 2 r r_g \cos(\alpha)}{2 r_i r_g}
  \label{eqn:APP_innertrans_lawcos}
\end{equation}

The law of sines yields an additional expression for $\alpha_0$ 

\begin{equation}
  \cos(\alpha_0) = \frac{r_p \sin(\phi)}{r_i}
  \label{eqn:APP_innertrans_lawsin}
\end{equation}

Eqs. (\ref{eqn:APP_innertrans_lawcos}) and (\ref{eqn:APP_innertrans_lawsin}) can be combined to yield an expression for $\alpha$

\begin{equation}
  \cos(\alpha) = \frac{2 r_p r_g \sin(\phi) - \left(r^2 - r_i^2 \right)}{2 r r_g}
  \label{eqn:APP_alpha_innertrans}
\end{equation}

As in the previous cases, expanding the phantom from a point to a small sphere of radius $R$ enables reexpressing Eq.\ (\ref{eqn:APP_alpha_innertrans})

\begin{equation}
  \cos(\alpha) = \frac{2 (r_p \pm R\csc(\phi)) r_g \sin(\phi) - \left(r^2 - r_i^2 \right)}{2 r r_g}
  \label{eqn:APP_alpha_innertrans_R}
\end{equation}

which yields

\begin{equation}
  \Delta\alpha = \frac{2R}{r}
  \label{eqn:APP_delta_alpha_innertrans}
\end{equation}

and

\begin{equation}
  \Delta\theta = \frac{4R}{\sqrt{4r^2 + (h_s \pm 2h_p)^2}}
  \label{eqn:APP_delta_theta_innertrans}
\end{equation}

where, as before, the $+$ case corresponds to the bottom surface and the $-$ case to the top. Once again, the solid angle element can be computed using $\Delta\alpha$ and $\Delta\theta$
\begin{samepage}
\begin{align}
  \Delta\Omega &= \frac{\pi}{4}\int_{\alpha-\Delta\alpha/2}^{\alpha+\Delta\alpha/2} \int_{\theta-\Delta\theta/2}^{\theta+\Delta\theta/2} \sin(\theta')\mathrm{d}\theta'\mathrm{d}\alpha' \label{eqn:APP_delta_Omega_innertrans}\\
  &= \frac{4 \pi R^2 }{\left(4r^2 + (h_s \pm 2h_p)^2\right)} \notag
\end{align}
\end{samepage}

and the projected area element is given by

\begin{samepage}
\begin{align}
  \mathrm{d}A_{\text{proj}} &= \cos(\theta)\sin(\alpha)\mathrm{d}A \label{eqn:APP_dA_proj_innertrans}\\
  &= \frac{h_s \pm 2h_p}{\sqrt{r^2 + (h_s \pm 2h_p)^2}} \notag \\
  & \hspace{1 cm}\mathrm{Re} \left[\sqrt{1 - \left(\frac{ r^2 + r_p^2 - {r_i}^2 - 2(r+\mathrm{r_g}(K,r,\phi,\mp \frac{h_s}{2}))r_p\cos(\phi)}{2 r \,\mathrm{r_g}(K,r,\phi, \mp \frac{h_s}{2})}\right)^2}\right] r \mathrm{d}\phi \mathrm{d}r \notag
\end{align}
\end{samepage}

Putting it all together, the flux through the inner transition surface is given by 

\begin{samepage}
\begin{align}
  \Phi_{\text{t,i}}(K) &= \int_{0}^{2\pi}\int_{\mathrm{r_i'}(r_p,\phi)}^{\mathrm{r_{\text{max}}}(K,\phi)}  \frac{4 r \left(h_s \pm 2h_p\right) }{\left(4r^2 + \left(h_s \pm 2h_p\right)^2\right)^{3/2}}  \label{eqn:APP_FI_trans_inner}  \\
  & \hspace{1 cm}\mathrm{Re} \left[\sqrt{1 - \left(\frac{ r^2 + r_p^2 - {r_i}^2 - 2(r+\mathrm{r_g}(K,r,\phi,\mp \frac{h_s}{2}))r_p\cos(\phi)}{2 r \,\mathrm{r_g}(K,r,\phi, \mp \frac{h_s}{2})}\right)^2}\right] \mathrm{d}r\mathrm{d}\phi \notag \notag \\
\end{align}
\end{samepage}

where the upper limits of integration in $r$ are given by 

\begin{multline}
  \mathrm{r_{\text{max}}}(K,\phi) = \text{clamp}\left( \frac{(r_o^2 - r_i^2)(h_s \pm 2h_p)}{4 \mathrm{r_g}(K,r,\phi,h_p)(r_o+r_p\sin(\phi))} \right. \\ \left. \mathrm{Re}\left[\sqrt[-2]{1 - \left( \frac{r_o^2-r_i^2}{2\mathrm{r_g}(K,r,\phi,h_p)(r_o+r_p\sin(\phi)) } \right)^2} \right], \mathrm{r_i'}(r_p,\phi), \mathrm{r_o'}(r_p,\phi) \right)
  \label{eqn:APP_rlim}
\end{multline}

\subsection{Endcap Surface}

The derivation of the expression describing GCR flux through the endcap surface largely follows that of the outer transition region, minus the magnetic attenuation component. The solid angle element for the endcap is given by Eq. (\ref{eqn:APP_delta_Omega_outtrans}), while the projected area element is given by

\begin{align}
  \mathrm{d}A_{\text{proj}} &= \cos(\theta)\sin(\alpha)\mathrm{d}A \label{eqn:APP_dA_proj_endcap} \\
  &= \frac{(h_s \pm 2h_p)}{\sqrt{4r^2 + (h_s \pm 2h_p)^2}} r \mathrm{d}\phi \mathrm{d}z \notag
\end{align}

Together, these yield the flux integral describing GCR flux through the endcap surface

\begin{equation}
  \Phi_{\text{e}}(K) = \int_{0}^{2\pi}\int_{0}^{\mathrm{r_i'}(r_p,\phi)}  \left( \frac{4 r \left(h_s-2h_p\right) }{\left(4r^2 + \left(h_s-2h_p\right)^2\right)^{3/2}} +\frac{4 r \left(h_s+2h_p\right) }{\left(4r^2 + \left(h_s+2h_p\right)^2\right)^{3/2}} \right) \mathrm{d}r \mathrm{d}\phi
  \label{eqn:APP_FI_endcap}
\end{equation}

All 5 flux integrals are then numerically integrated in MATLAB to produce the resulting GCR flux incident upon the spherical phantom. This attenuated spectrum is then passed to HZETRN2020, which handles the nuclear physics and interactions and energy deposition calculations.

\section{Monte Carlo Validation}

\renewcommand{\theequation}{C.\arabic{equation}}
\renewcommand{\thefigure}{C\arabic{figure}} 
\renewcommand{\thetable}{C\arabic{table}} 
\setcounter{equation}{0}
\setcounter{figure}{0}

The Monte Carlo simulation used to validate the new magnetic attenuation model is further discussed here. A GCR proton is initialized at random point on a sphere of radius $r_{\text{inf}} =\text{100 m}$, much greater than the longest dimension of the shield. A spherical cap is then generated with pole aligned with the $-\hat{r}$ (i.e., radial) direction and half-angle given by 

\begin{equation}
  D = \arctan(\frac{\max(h_s/2,r_o)}{r_{\text{inf}}})
  \label{eqn:APP_MCcap}
\end{equation}

The proton's initial velocity is then randomly rotated under the constraint that it is directed towards a point along this spherical cap. Setting $D$ in the manner above ensures the particle flux upon the shield is entirely isotropic, minimizing any biases that may be introduced via other methods, while minimizing the number of generated particles that miss the shield entirely. 

The particle is then evolved forwards in time, travelling towards the shield. When the particle is outside the shield annular cylinder, it is assumed that no other forces act upon it and so continues to travel in a straight line. Therefore, the particle can be immediately moved to the point where it initially intersects the shield (or the point where it exits the simulation domain, should it miss the shield entirely). Once inside the shield annular cylinder, the particle is evolved using the pure-B Boris algorithm \citep{boris_relativistic_1971}, wherein the particle is rotated according to the influence of the magnetic field, accounting for relativistic effects

\begin{equation}
  \vect{p}_{i+1} = \vect{p}_{i} + (\vect{p}_{i} \times \vect{T}) \times \vect{S}
  \label{eqn:APP_BorisPush}
\end{equation}

where $T$ and $S$ are given by 

\begin{eqnarray}
  \vect{T} &=& \frac{q \,  \Delta t}{2 m \, \gamma_i} \vect{B} \\
  \vect{S} &=& \frac{2 \, \vect{T}}{1 + \vect{T} \cdot \vect{T}}
  \label{eqn:APP_BorisTS}
\end{eqnarray}

where the timestep $\Delta t$ is set to $0.223/V$.

By applying a rotation to the relativistic momentum, the Boris algorithm preserves the magnitude of the momentum much more accurately than traditional solution methods (e.g., Runge-Kutta). Traditional methods include associated truncation errors that accumulate and will significantly alter the long-term trajectories of relativistic particles. 

This process is process is repeated to create a set of incident particles that, when a sufficiently large number of particles are generated, adequately represent the isotropic GCR environment encountered in deep space. The proportion of particles striking a spherical phantom at each energy is recorded. To enable comparison with the analytical methods, this proportion is then normalized by the proportion of particles that strike the phantom when the shield is removed (approximately 0.0035 for the process described in this section).


\clearpage

\bibliographystyle{cas-model2-names}

\bibliography{RadiationShielding}



\end{document}